\documentstyle[prc,aps,epsfig]{revtex}
\newcommand{\gsim}{\mathrel{\hbox{\rlap{\lower.55ex \hbox {$\sim$}}
                   \kern-.3em \raise.4ex \hbox{$>$}}}}
\newcommand{\lsim}{\mathrel{\hbox{\rlap{\lower.55ex \hbox {$\sim$}}
                   \kern-.3em \raise.4ex \hbox{$<$}}}}
\newcommand{\be}{\begin{equation}}
\newcommand{\ee}{\end{equation}}
\newcommand{\ba}{\begin{eqnarray}}
\newcommand{\ea}{\end{eqnarray}}
\newcommand{\no}{\nonumber \\}
\newcommand{\etal}{{\it et al.}~}
\newcommand{\eg}{{\it e.g.}}
\newcommand{\ie}{{\it i.e.}}

\def\Tr{\rm Tr}

\begin{document}

\tightenlines

\title{A Schematic Model for $\rho$-$a_1$ Mixing at Finite
Density and In-Medium Effective Lagrangian}

\author{Youngman Kim\footnote{Current address: 
Department of Physics, Hanyang University, Seoul 133-791,
Korea}, R. Rapp, G.E. Brown 
and Mannque Rho\footnote{Permanent address: 
Service de Physique Th\'eorique, CE Saclay 91191
Gif-sur-Yvette, France} }

\address{Department of Physics and Astronomy, State
University of New York Stony Brook, NY 11794-3800, USA}

\maketitle

\begin{abstract}
 
 Based on schematic two-level models
extended to $a_1$-meson degrees of freedom, we investigate possible
mechanisms of chiral restoration in the vector/axialvector
channels in cold nuclear matter. In the first part of this article
we employ the massive Yang-Mills framework to construct an
effective chiral Lagrangian based on low-energy mesonic modes at
finite density. The latter are identified through nuclear
collective excitations of `meson'-sobar type such as
$\pi\leftrightarrow [\Delta (1232)N^{-1}]\equiv\hat\pi$,
$\rho\leftrightarrow [N^* (1520)N^{-1}]\equiv\hat\rho$, etc..
In a mean-field type treatment the in-medium gauge coupling $\hat
g$, the (axial-) vector meson masses and $\hat f_\pi$ are found to
decrease with density indicating the approach towards chiral
restoration in the language of in-medium effective fields.
In the second part of our analysis we evaluate the (first)
in-medium Weinberg sum rule which relates vector and axialvector
correlators to the pion decay constant. Using in-medium
$\rho$/$a_1$ spectral functions (computed in the two-level model)
also leads to a substantial reduction of the pion decay constant
with increasing density.

\end{abstract}
 
\pacs{25.75.-q, 21.65.+f, 12.39.Fe}

%%%%%%%%%%%%%%%%%%%%%%%%%%%%%%%%%%%%%%%%%%%%%%%%%%%%%%%%%%%%%%%%%%%%
\section{Introduction}
\label{sec_intro}
%%%%%%%%%%%%%%%%%%%%%%%%%%%%%%%%%%%%%%%%%%%%%%%%%%%%%%%%%%%%%%%%%%%%

The density dependence of vector-meson masses encoded in the
so-called Brown/Rho (BR) scaling conjecture~\cite{br} has stimulated
considerable discussion. From the experimental side, the CERES
dilepton experiments~\cite{ceres} have provided strong evidence
that the properties of $\rho$ mesons are substantially modified in
hot/dense matter. The measurements performed at the full SpS
energy indicate an excess of dileptons with invariant masses below
$\sim$~0.6~GeV, as well as missing strength in the region around
the free $\rho$-mass. More quantitative results are expected from
further runs at both 40 and 158~AGeV with an additional TPC improving 
the mass resolution in order to discriminate the contributions of
final state $\omega$ decays from the $\rho$-meson decays within
the interacting hadronic fireball.

The simplest and most economical explanation for the observed
low-mass dilepton spectra is achieved in terms of quasiparticles
(both fermions and bosons) whose masses drop according to BR
scaling, thereby making an appealing link to the chiral (quark-)
structure of the hadronic vacuum. In an alternative view to this
description, Rapp, Chanfray and Wambach (R/W)~\cite{RCW} showed
that the excess of low-mass dileptons follows from
hadronic many-body calculations using in-medium spectral functions
(see, \eg, Ref.~\cite{RW00} for a recent review). On rather
general grounds, this ``alternative'' description was in a sense
anticipated as discussed by one of the authors~\cite{duality}. In
analogy to the quark-hadron duality in heavy-light meson decay
processes, one may view BR scaling as a ``partonic'' picture
while R/W as a hadronic one. Loosely speaking, on the finite
density axis, the former can be thought of as a top-down approach
and the latter as a bottom-up one. The link between BR scaling and
the Landau quasiparticle interaction $F_1$ established in
\cite{FR96} is one specific indication for this ``duality''.
Indeed, in \cite{krbr}, Brown~\etal argued that the R/W
explanation encodes features of a density-dependent $\rho$-meson
mass, calculated in a hadronic basis (in contrast to that of
constituent quarks used by Brown and Rho). In particular it was
suggested~\cite{krbr} that if one replaces the $\rho$-meson mass
$m_\rho$ by the mass $m_\rho^*(\rho)$\footnote{We will discuss the 
reasoning behind this suggestion in Sect.~\ref{sec_schem}.} at the 
density being considered, one would arrive at a description, in hadron
language, which at high densities appeared dual to that of the
Brown/Rho one in terms of constituent quarks. These developments
involved the interpretation of a {\it collective} isobar-hole
excitation as an effective vector-meson field operating on the
nuclear ground state, {\it i.e.},
\be
\frac{1}{\sqrt{A}}\sum_{i}[N^*(1520)_iN_i^{-1}]^{1-}
\simeq \sum_{i} [\rho(x_i)~{\rm or}~ \omega(x_i) ]|\Psi_0>_s \ ,
\label{col}
\ee
with the antisymmetric (symmetric) sum over neutrons and
protons giving the $\rho$-like ($\omega$-like) nuclear excitation.
The dropping vector meson masses could then be estimated in terms
of the mixing of this collective state with the
elementary vector meson state~\cite{krbr}.

In the present work, we will expand on these ideas by constructing
an effective chiral Lagrangian involving effective fields for
`meson-sobars' ($\hat\pi ,\hat\rho$, ...)  which in a dense medium 
are assumed to be the relevant, lowest-lying degrees of freedom
in terms of the nuclear collective states in the  corresponding
meson channels. Therefore we will assume that each meson field has
its `sobar' partner,\footnote{In general, the sobar field would be
a linear combination of $N^*$-hole states of the appropriate
quantum numbers but here, for simplicity, we are taking only what
we consider to be the dominant component.}  that is
$\pi\leftrightarrow [\Delta (1232)N^{-1}]\equiv\hat\pi$,
$\rho\leftrightarrow [N^*(1520)N^{-1}]\equiv\hat\rho$ and
$a_1\leftrightarrow [N^*(1900)N^{-1}]\equiv\hat{a_1}$. We then
construct a  chiral effective Lagrangian for meson-sobar fields
following the procedure of the massive Yang-Mills
approach~\cite{mym}. An explicit $a_1$-meson is not necessarily
required to formulate a chiral invariant Lagrangian involving
$\rho$-mesons, as is well-known, \eg, from the hidden local
symmetry framework~\cite{hls}; in our framework, however, it
provides a convenient  treatment of  the associated low-lying mode
on an equal footing with the $\rho$-meson thereby facilitating the
discussion of chiral restoration in the vector-axialvector
doublet. Moreover, in the mean-field analysis carried out below we
find that at some density our meson-sobar fields could be
described in terms of interpolating fields that are the effective
fields figuring in an in-medium Lagrangian exhibiting BR scaling.
One of the main differences between the fields of BR scaling and
meson-sobar fields is that while in the former the full pole
strength ($Z\sim 1$) is retained by the low-lying mode, the sobar
fields only carry a fraction of the strength in the respective
meson channel (typically $Z\sim 0.3$). It will be suggested that
this discrepancy can be resolved by applying a (finite)
wave-function renormalization to the  sobar fields.

Our article is organized as follows:
in Sect.~\ref{sec_schem} we review the schematic model~\cite{krbr}
and extend it by including $a_1$ meson. The corresponding meson-sobar
chiral Lagrangian and the ensuing finite-density results in
mean-field approximation are presented in Sect.~\ref{sec_sobar}.
Using the same schematic model from Sect.~\ref{sec_schem}, but following the
philosophy of the many-body spectral function approach~\cite{RCW,RW99}, we
compute in Sect.~\ref{sec_wsr} the density dependence of the pion decay
constant employing the in-medium Weinberg sum rules.
Sect.~\ref{sec_concl} contains a summary and concluding remarks.

%%%%%%%%%%%%%%%%%%%%%%%%%%%%%%%%%%%%%%%%%%%%%%%%%%%%%%%%%%%%%%%%%%%%%%%
\section{The Schematic Model}
\label{sec_schem}
%%%%%%%%%%%%%%%%%%%%%%%%%%%%%%%%%%%%%%%%%%%%%%%%%%%%%%%%%%%%%%%%%%%%%%%

Let us first briefly review the main features of the schematic
model for the in-medium $\rho$-meson as used in \cite{krbr} and
then extend it to the $a_1$ channel.

The starting point is the $\rho$-meson propagator in nuclear matter given by
\be
D_\rho(q_0,\vec q)=\frac{1}{q_0^2-\vec q^2 -m_\rho^{2}+im_\rho
\Gamma_{\pi\pi}(M)-\Sigma_{\rho N^* N}(q_0,q)} \ ,  
\label{rhop1} 
\ee
where $\Gamma_{\pi\pi}$ denotes the vacuum $\rho$ decay width, and
the real part of the selfenergy has been absorbed into the free
(physical) mass $m_\rho$. The entire density dependence resides in
the in-medium $\rho$ selfenergy $\Sigma_{\rho N^* N}$ induced by
$N^*(1520)N^{-1}$ excitations. It is calculated from the
interaction Lagrangian
\be
{\cal L}_{\rho NN^*(1520)}=
\frac{f_\rho}{m_{\rho}}\psi_{N^*}^\dagger (q_0\vec S\cdot\vec
\rho_a - \rho_a^0\vec S\cdot \vec q) \tau_a \psi_N  + ~  h.c. \  ,
\label{L1520} 
\ee 
where the coupling constant can be estimated
from the measured branching of the $N^*(1520)\to \rho N$ decay (as
well as information on the radiative decay~\cite{RUBW}). In what
follows we will for simplicity concentrate on the limit of
vanishing three-momentum where the longitudinal and transverse
polarization components become  identical. Due to the rather high
excitation energy of $\Delta E=M_{N^*}-M_N=580$~MeV, one can
safely neglect nuclear Fermi motion to obtain
\be
\Sigma_{\rho N^* N}(q_0)= \frac{8}{3} f_{\rho N^* N}^2
\frac{q_0^2}{m_\rho^2}\frac{\rho_N}{4} \left(\frac{2(\Delta
E)}{(q_0+i\Gamma_{tot}/2)^2-(\Delta E)^2}\right) \label{RWself}
\ee ($\rho_N$: nucleon density). If the widths of the $\rho$ and
$N^*(1520)$ are sufficiently small one can invoke the mean-field
approximation (as employed in Sect.~\ref{sec_sobar}) and determine the
quasiparticle excitation energies from the zeros in the real part
of the inverse propagator. In particular, for $\vec q=0$ the
in-medium $\rho$ mass is obtained by solving the dispersion
relation \ba q_0^2 =m_\rho^2+{\rm Re} \Sigma_{\rho N^* N}(q_0) \ .
\label{disp} \ea The pertinent spectral weights of the solutions
are characterized by $Z$-factors defined through
\be
Z=(1-\frac{\partial}{\partial q_0^2}{\rm Re} \Sigma_{\rho N^* N})^{-1} \ .
\label{Zfac}
\ee

Within a chirally invariant framework, we need to include the
chiral partners of the $\rho$ and its nuclear excitation, \ie, the
$a_1$ and a suitable $N^*$ resonance with spin-3/2 and positive
parity. A possible candidate is the $N^*(1900)$ state, and the
interaction Lagrangian is taken in analogy to (\ref{L1520})
as\footnote{We could also consider coupling terms of the type
$a_1N^*(1520)N$ as well as $\rho N^*(1900)N$, which in their
relativistic version involve an additional $\gamma_5$ as compared
to the ones used here; in the nonrelativistic reduction this leads
to selfenergies of $P$-wave nature being  proportional to  $\vec
q$ and can therefore be neglected in the zero-momentum limit
considered here.}
\be
{\cal L}_{a_1 NN(1900)}
=\frac{f_{a_1}}{m_{a_1}}\psi_{N^*}^\dagger (q_0\vec S\cdot\vec A_a -
A_a^0\vec S\cdot \vec q)t_a\psi_N ~+~h.c. \
\ee
with $A$ denoting the axialvector $a_1$-field.
The coupling constant $f_{a_1}$ can in principle be estimated from the
the partial decay width $\Gamma_{N^*(1900)\rightarrow a_1 N}$. Although
the corresponding 3-pion final state has not been explicitely measured,
the observed~\cite{pdg,Man92}
 one- and two-pion final states leave room for up to 30\%
branching of the total $N^*(1900)$ width of $\sim$~500~MeV into
the $a_1N$ channel. Using the baryon decay width formula from Ref.
\cite{RW00}, \ba \Gamma_{N(1900)\rightarrow a_1
N}(\sqrt{s})&=&\frac{f_{a_1}^2}{4\pi m_{a_1}^2}
\frac{2m_N}{\sqrt{s}} \frac{2I_a+1}{(2J_B+1)(2I_B+1)} SI(a_1 NN^*)
\no && \times \int_{3m_\pi}^{\sqrt{s}-m_N}\frac{M}{\pi}dMA_a^0(M)
q_{cm}F_{ANB}(q_{cm}^2)^2(M^2+2q_0^2) \no
 \ea
with a standard monopole form factor 
$F_{a_1NB}(q_{cm}^2)= \Lambda_{a_1}^2/(\Lambda_{a_1}^2+q_{cm}^2)$
($\Lambda_{a_1}=600$~MeV as for the $\rho$), we find $f_{a_1}\simeq
25.3$ for $20 \%$ and $f_{a_1}\simeq 17.8$ for $10 \% $ branching
ratio. This range of values will be used to indicate the inherent
uncertainties in our numerical results presented below.

\begin{figure}[htb]
\epsfig{file=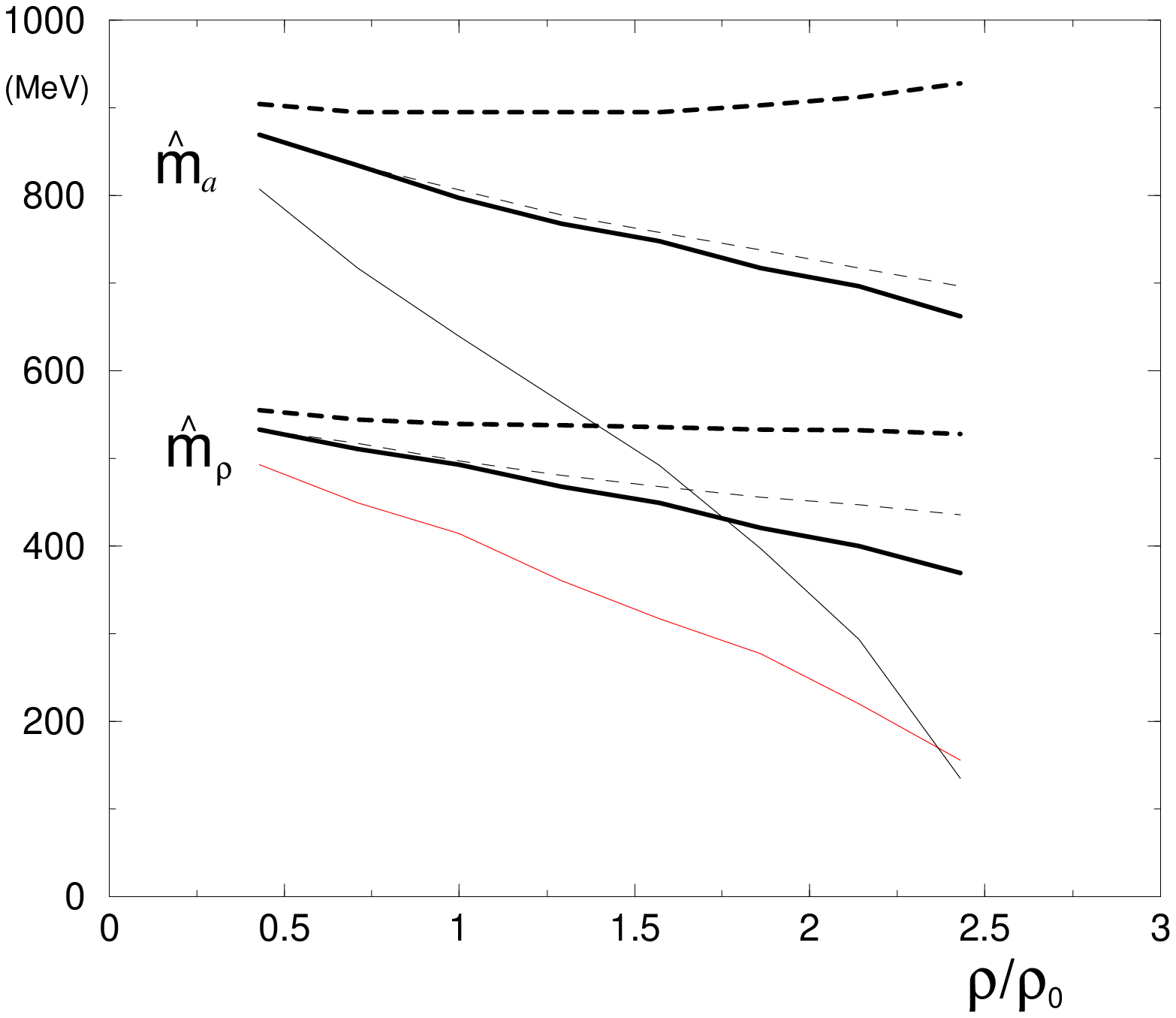,width=8cm}
\hspace{1cm}
\epsfig{file=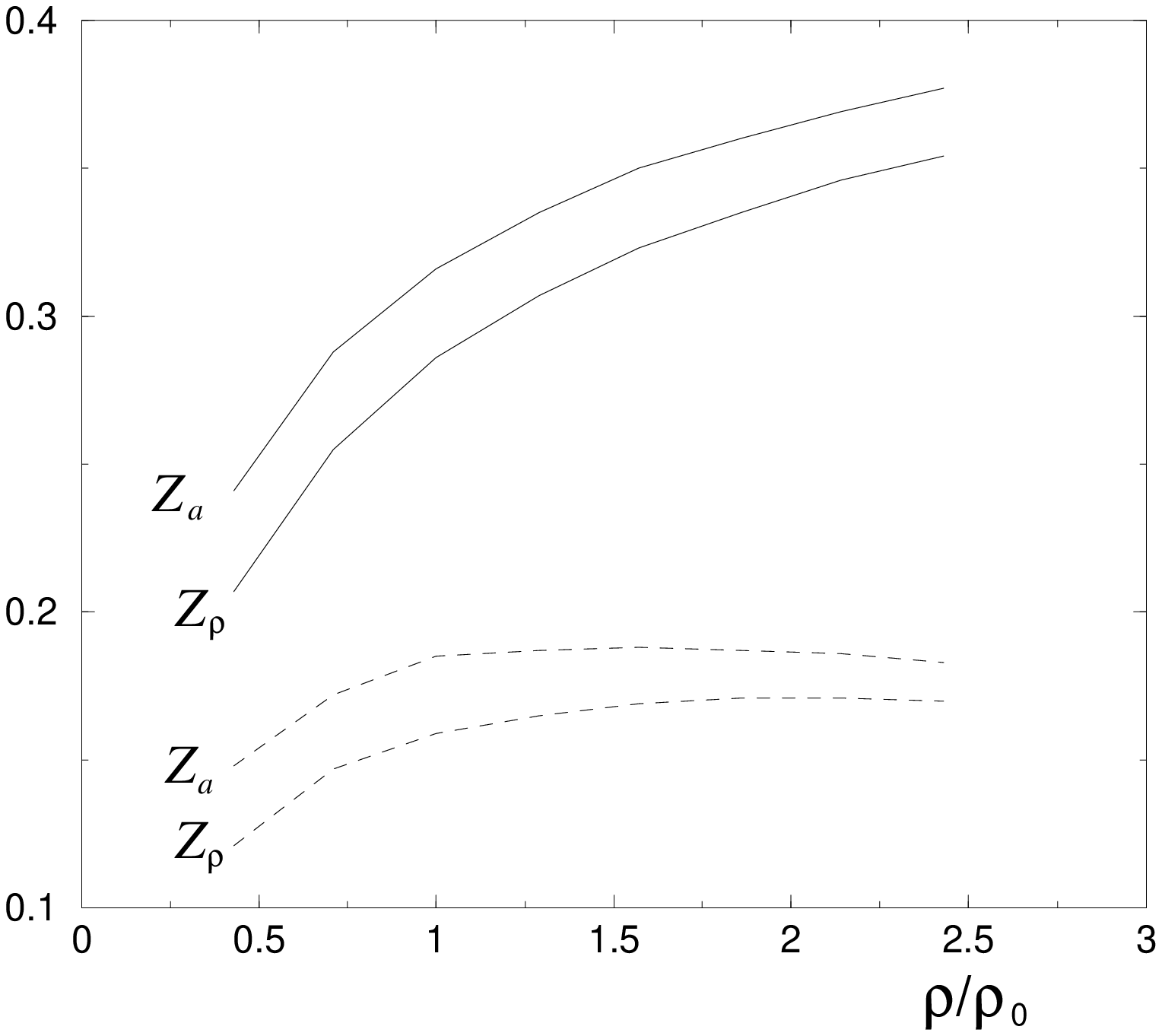,width=7.8cm}
\vspace{0.3cm}
\caption{\small Left panel: nn-medium rhosobar and $a_1$-sobar masses in the
R/W (dashed lines) and BR (solid lines) approach.
Thick lines represent in-medium masses with $\Gamma_{tot}=\Gamma_{tot}(\rho)$
and thin lines are for $\Gamma_{tot}=0$. 
Right panel: $Z$-factors for  rhosobar ($Z_\rho$)
and $a_1$-sobar ($Z_a$) in the R/W (dashed lines) and BR (solid lines)
approach for $\Gamma_{tot}=0$}
\label{fig_schem}
\end{figure}

Before we address the construction of the full chiral Lagrangian
in the next section, we compute the density dependence of the
masses and $Z$-factors corresponding to the low-lying $\rho$- and
$a_1$-sobar states when selfconsistently solving the dispersion
relation (\ref{disp}). Using the selfenergy as given in
eq.~(\ref{RWself}) (and the analogous expression for the
$a_1$-sobar) results are depicted by the dashed lines in 
Fig.~\ref{fig_schem} (corresponding to the R/W approach in
Ref.~\cite{krbr}). When neglecting the baryon resonance-widths,
one finds a moderate simultaneous decrease of both $\hat{m_\rho}$
and $\hat{m_{a_1}}$ associated with pole strengths of 10-20\%.
Upon inclusion of resonance widths $\Gamma_{tot}=\Gamma_0
+\Gamma_{med}$ (with vacuum values $\Gamma_0=120 (500)$~MeV for
$N^*(1520)$ ($N^*(1900)$) and medium corrections
$\Gamma_{med}\simeq 300\rho_N/\rho_0$~MeV as inferred in
Refs.~\cite{peters,RUBW}) the collectivity is suppressed and
little density dependence is observed. However, since large widths
also imply that the quasiparticle (mean-field) approximation
becomes less reliable, it would be premature to conclude from the
behavior of the masses alone that there is no  approach towards
chiral restoration. We will come back to this issue in Sect.~\ref{sec_wsr}.

The situation quantitatively changes if one adopts the suggestion put
forward in Ref.~\cite{krbr} to replace in the selfenergy expressions
the $1/m_\rho^2$ ($1/m_{a_1}^2$) factor by the (selfconsistent)
in-medium mass $1/(m_\rho^*)^2$ ($1/(m_{a_1}^*)^2$) (B/R approach;
full lines in Fig.~\ref{fig_schem})\footnote{We
have no compelling argument for the validity of this procedure. It
is based on the following conjecture, the proof of which would
require going beyond the framework we are developing here. The
free $\rho N^* N$ coupling is of the form $\frac{f}{m}q_0$ with a
dimensionless constant $f$ and $q_0$ the fourth component of the
the $\rho$ meson momentum. Writing this as $Fq_0$ with $F=f/m$ the
medium renormalization of the constant $F$ will certainly depend
on density $\rho$. In order for the vector meson mass to go to
zero at some high density so as to match B/R scaling, it is
required that $F(\rho)q_0 \rightarrow {\rm constant}\neq 0$. For
$q=|\vec{q}|\approx 0$ which we are considering, this can be
satisfied if $F(\rho)\sim {m^*}^{-1}$, modulo an overall constant.
This is the essence of the proposal of Ref.~\cite{krbr}.}. The
dispersion relation then takes the form
\be
q_0^2 =m_\rho^2+f_{\rho N^* N}^2\frac{4}{3}\rho_N \frac{\Delta
E(q_0^2-(\Delta E)^2-\frac{1}{4}\Gamma_{tot}^2)} {(q_0^2-(\Delta
E)^2-\frac{1}{4}\Gamma_{tot}^2)^2+\Gamma_{tot}^2q_0^2} \ .
\label{dis1} 
\ee 
Here, we shall assume that in the medium $\Delta E$ remains 
unchanged (which can be argued to hold to the leading
order in $1/N_c$) while the width $\Gamma_{tot}$ may be affected
by density (as before). With  $f_{a_1} \simeq 17.8$ (corresponding
to $10 \%$ branching ratio) the mean-field results
($\Gamma_{tot}=0$) lead to  chiral symmetry restoration  around
$\rho =2.4\rho_0$; again, inclusion of finite widths decelerates
the drop in masses in this schematic treatment (in the more
extreme case with $f_{a_1} \simeq 25.3$ the in-medium $a_1$-meson
mass drops to zero around $\rho\sim 1.2\rho_0$ for $\Gamma_{tot}
=0$). The `B/R approach' in evaluating the $\rho$- and
$a_1$-sobar selfenergies is not based on rigorous arguments, but it
successfully reproduces BR scaling leading to chiral restoration
through $\hat m_\rho, \hat m_{a_1}$ merging together
close to zero somewhat below 3 times nuclear matter density.

%%%%%%%%%%%%%%%%%%%%%%%%%%%%%%%%%%%%%%%%%%%%%%%%%%%%%%%%%%%%%%%%%%%
\section{Effective Lagrangian for Meson-Sobars}
\label{sec_sobar}
%%%%%%%%%%%%%%%%%%%%%%%%%%%%%%%%%%%%%%%%%%%%%%%%%%%%%%%%%%%%%%%%%%%

%%%%%%%%%%%%%%%%%%%%%%%%%%%%%%%%%%%%%%%%%%%%%%%%%%%%%%%%%%%%%%%%%%%
\subsection{Mean-Field Model and in-Medium Pion Decay Constant}
%%%%%%%%%%%%%%%%%%%%%%%%%%%%%%%%%%%%%%%%%%%%%%%%%%%%%%%%%%%%%%%%%%%

Our subsequent analysis in this section will be based on the
following two main assumptions: (i) each meson field has its
'sobar'-partner, \ie, $\pi\leftrightarrow [\Delta
(1232)N^{-1}]\equiv\hat\pi$, $\rho\leftrightarrow [N^*
(1520)N^{-1}]\equiv\hat\rho$ and $a_1\leftrightarrow
[N(1900)N^{-1}]\equiv\hat a_1$, and, (ii)  chiral symmetry
persists in the meson-sobar subspace. Before spelling out the
chiral Lagrangian, let us elaborate on the pertinence of these
propositions\footnote{To see how to identify, \eg, the $\hat\pi$
field, consider the equation of motion for pions with ${\cal
L}_{\pi N\Delta}$ in (\ref{ints}), $(\partial_\mu\partial^\mu
-m_\pi^2)\vec\pi =\frac{f_{\pi N\Delta}^*}{m_\pi}\partial^\mu
(\bar\psi_{\Delta,\mu}\vec T\psi_N ) +h.c. $ with
$\psi_{\Delta,\mu}$ denoting the $\Delta$-field. Taking the
expectation value and ignoring the kinetic term, we have $<\vec\pi
>= \frac{1}{m_\pi^2} <\frac{f_{\pi N\Delta}^*}{m_\pi}\partial^\mu
(\bar\psi_{\Delta,\mu}\vec T \psi_N ) +h.c.>$. Thus the
$\pi$-sobar field can be identified as
 \ba \hat{\vec\pi}
=\frac{1}{m_\pi^2} <\frac{f_{\pi N\Delta}^*}{m_\pi}\partial^\mu
(\bar\psi_{\Delta,\mu}\vec T\psi_N ) +h.c.>. \nonumber
 \ea}.
In the context of our two-level schematic 
model in Sect.~\ref{sec_schem}, a meson
Green's function, $<\phi\phi^+>$, can develop a new pole (i.e.,
excitation), $<\hat\phi\hat\phi^+>$ with associated pole strength
$Z$, due to the effects of a many-particle medium; at the same
time, the  original (`elementary') excitation, $<\phi_e\phi_e^+>$,
will be shifted from its original  pole position and carry the
residue $1-Z<1$. There appears no mixed Green's function,
$<\phi_e\hat\phi^+ + h.c.>$ since the mixing matrix of the nuclear
collective state (resonance-hole state $ \approx$ $\hat\phi$) and
the elementary meson ($\approx \phi_e$) is diagonalized. This
implies that \ba {\cal L}(\phi )\approx {\cal L}(\phi_e )+ {\cal
L}(\hat\phi )+ {\cal L}(\hat\phi,\phi_e ) \approx
 {\cal L}(\phi_e )+ {\cal L }(\hat\phi )
\ea
as far as kinetic and mass terms are concerned, \ie,
if the original Lagrangian $L(\phi)$ possesses a (linearly realized)
chiral symmetry, then the non-interacting part of
${\cal L}(\hat\phi )$ preserves it.
For the interaction parts, as well as for
a  nonlinear realization of chiral symmetry, it is not so clear how to
demonstrate the existence of chiral symmetry in our meson-sobar Lagrangian
using the arguments given above.

Employing the Massive Yang-Mills (MYM) framework~\cite{mym},
we can write a chiral effective Lagrangian
with $SU(2)_L\times SU(2)_R$ symmetry for meson-sobar fields
with minimal couplings,
\ba
{\cal L}&=& \frac{1}{4}\hat f_\pi^2 {\Tr} (D_\mu\hat U D^\mu\hat U^\dagger )
+ \frac{1}{4} \hat f_\pi^2 {\Tr} [ M(\hat U+\hat U^\dagger -2)]
-\frac{1}{2}{\Tr} (\hat F_{\mu\nu L}\hat F_L^{\mu\nu} 
+\hat F_{\mu\nu R}\hat F_R^{\mu\nu})
\no
&&+\hat m_0^2(\hat A_{\mu L}\hat A_L^\mu + \hat A_{\mu R} \hat A_R^\mu) \ , 
\label{cel-1}
\ea
where
\ba
\hat U &=& e^{\frac{2i\hat\pi}{\hat f_\pi}}~,
\ \hat\pi=\hat\pi^a\frac{\tau^a}{2}\no
\hat F_{\mu\nu}^{L,R}&=&\partial_\mu \hat A_\nu^{L,R} -
\partial_\nu \hat A_\mu^{L,R}
(\hat V_\mu \pm \hat A_\mu) \no
D_\mu \hat U &=&\partial_\mu \hat U -igA_{\mu L} \hat U +igA_{\mu R},\no
M &=& m_\pi^2 {\bf 1} \ .
\label{la1}
\ea
Note that we have the same mass matrix $M$ as appearing in Ref.~\cite{mym}.
It remains to fix the free parameters in (\ref{la1}) in terms of
$f_\pi$, $g$ and the masses of the original MYM-Lagrangian~\cite{mym}.

\begin{figure}[htb]
\centerline{\epsfig{file=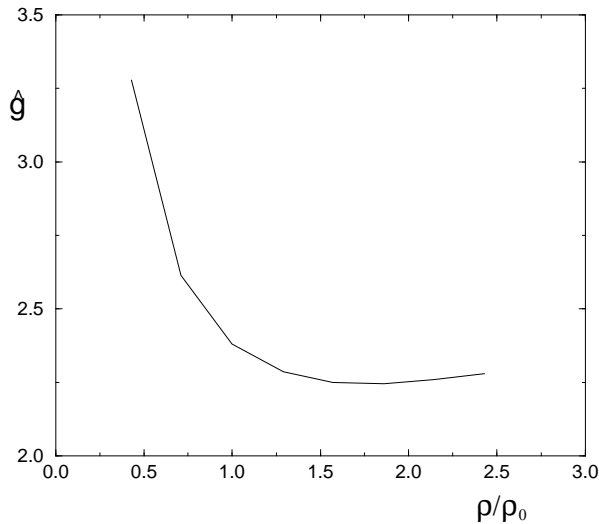,width=8cm}}
\vspace{0.3cm}
\caption{\small Density dependence of $\hat g$.}
\label{ghat}
\end{figure}

Following the procedure in \cite{mym}, the physical mass
difference between $\hat a_1$  and $\hat\rho$ is obtained after
diagonalizing the quadratic piece of the Lagrangian (\ref{cel-1})
to remove the spurious $A_\mu \partial^\mu \pi$ mixing 
term,\footnote{In the following our parameters and fields will always
correspond to the physical ones, denoted by  $\tilde f_\pi$,
$\tilde \pi$ and $\tilde A_\mu$ in Ref.~\cite{mym}.} 
\ba 
\hat m_{a_1}^2 -\hat m_\rho^2 =\frac{\hat g^2\hat f_\pi^2}{2}
\frac{\hat m_{a_1}^2}{\hat m_\rho^2}  \ , 
\label{mass2} 
\ea 
where $\hat m_\rho=\Delta E=1520-940=580$~MeV and $\hat
m_{a_1}=1900-940=960$~MeV. Rewriting (\ref{mass2}) as 
\ba 
\hat
f_\pi^2 &=& \frac{2}{\hat g^2 }\frac{\hat m_\rho^2}{\hat
m_{a_1}^2} (\hat m_{a_1}^2 -\hat m_\rho^2 ) \ , 
\label{relation}
\ea 
we notice that the knowledge of  the (in-medium) values of
$\hat g$, $\hat m_\rho$ and $\hat m_{a_1}$ allows to infer the
density dependence of $\hat f_\pi$.
\begin{figure}[htb]
\centerline{\epsfig{file=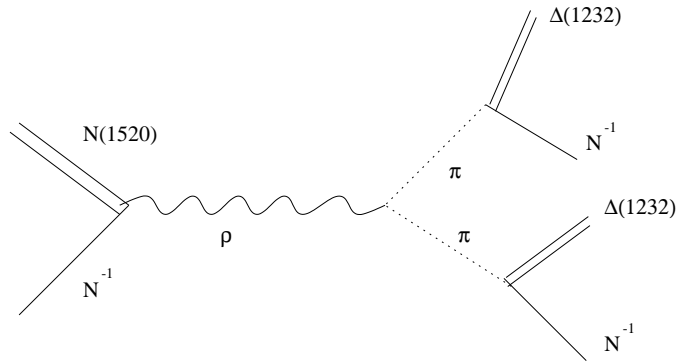,width=9cm}}
\vspace{0.3cm}
\caption{\small The diagram for the
decay process $\hat\rho \to \hat\pi \hat\pi$ of the low-lying
sobar-modes determining the effective
decay constant $\hat g_{\rho\pi\pi}$.}
\label{hrpp}
\end{figure}
The masses will be taken from the lower nuclear branches of the
schematic two-level model discussed in Sect.~\ref{sec_schem}. Note that 
the additional meson interaction vertices do not induce further medium
dependences at zero temperature. To estimate the value of the gauge
coupling constant $\hat g$, we use the relation $\hat
g_{\rho\pi\pi}=\frac{3}{4\sqrt{2}}\hat g$\cite{mym}, and assess
$\hat g_{\rho\pi\pi}$ by evaluating the diagram shown in
Fig.~\ref{hrpp}. With the standard phenomenological interaction
Lagrangians, 
\ba 
{\cal L}_{\pi N\Delta} &=& \frac{f_{\pi
N\Delta}}{m_\pi}\bar\psi_{\Delta,\mu} \vec
T\psi_N\cdot\partial^\mu\vec\pi \ + \ h.c.  
\no 
{\cal L}_{\rho N N^*} &=& \frac{f_{\rho NN^*}}{m_\rho}\bar\psi_{N^*,\nu}
 \gamma_\mu \vec \tau \psi_N \cdot (\partial^\mu\vec\rho^\nu
- \partial^\nu\vec\rho^\mu) \ + \ h.c. 
\no 
{\cal L}_{\rho\pi\pi} &=& g_{\rho\pi\pi} \vec\rho_\mu \cdot
(\vec\pi\times\partial_\mu\vec\pi) \ , 
\label{ints} 
\ea 
we obtain
in the $\rho$-meson rest frame, where $q^\mu=(\Delta E, \vec 0)$,
\ba 
{\cal L}_{\hat\rho\hat\pi\hat\pi}= \hat
g_{\rho\pi\pi}\hat{\vec\rho}_i \cdot\partial^i\hat{\vec\pi}\times
\hat{\vec\pi} \ 
\ea 
with\footnote{Within the spirit of the mean-field description 
pursued here we neglect in our estimate further vertex corrections
of, \eg, the $\rho\pi\pi$ coupling which are in principal
necessary to ensure gauge invariance of the vector current.} 
\ba 
\hat g_{\rho\pi\pi}= g_{\rho\pi\pi}
\frac{1}{\Delta E^2-m_\rho^2} \left( \frac{1}{(\Delta E /2)^2
-\vec k^2-m_\pi^2}\right)^2 \ m_\rho^2 m_\pi^4 \ 
\label{ghe} 
\ea
and $\partial^i$ corresponding to the spatial
components\footnote{Since Lorentz invariance is broken in the
medium, we will encounter non-covariant terms in our effective
Lagrangian. Our $\hat g$ and $\hat f_\pi$ may thus be considered
as the space components of the in-medium gauge coupling and pion
decay constant.}. 
A complication arises from the fact that $\hat g_{\rho\pi\pi}$ 
depends on the momentum of the pions and, consequently,
so will $\hat f_\pi$. Even worse, as written in eq.~(\ref{ghe}),
$\hat g_{\rho\pi\pi}$ diverges for a single-pion energy of
$\omega_k=\Delta E/2$, which essentially coincides with typical
$\pi$-sobar energies. The latter problem can be remedied by
accounting for (in-medium) widths for both pion and
rho propagators. Within the isobar-hole model the former takes
the standard form 
\ba 
D_\pi(\omega=\Delta E/2,\vec k) = \frac{1} {(\Delta E/2)^2-m_\pi^2-
\vec k^2 \ [1+ \chi_0 /(1-g^\prime\chi_0)]} \ , 
\ea 
where the pion susceptibility
$\chi$ is given by\cite{fm} 
\ba 
\chi (\omega ,\vec k)&\simeq&
\frac{8}{9}(\frac{f_\pi^*}{m_\pi})^2\frac{\Delta E_\pi +\vec
k^2/2m_\Delta} {(\omega +i\frac{1}{2}\Gamma_\Delta)^2 -(\Delta E_\pi 
+\vec k^2/2m_\Delta)^2} \ F_\pi(\vec k) \ \rho_N 
\ea 
with $\Delta E_\pi=M_\Delta-M_N\simeq 300$~MeV and a hadronic form
factor $F_\pi$. For the $\rho$-meson, we have 
\ba 
D_\rho= \frac{1}
{(\Delta E)^2-m_\rho^2-\Sigma_{\rho N^* N}(\Delta E)
+im_\rho\Gamma_\rho } \label{self2} 
\ea 
as before. 
At given density, we then average the momentum dependence of $\hat g$ over
an appropriate range according to $\hat g =\int_0^{k_c}
g(k)dk/k_c$ with $k_c\simeq 700$~MeV.
The final result for the density dependence of $\hat g$ is displayed
in Fig.~\ref{ghat} where we have neglected any medium dependencies
of $m_\rho$, $\Delta E$, $m_\pi$, $\Delta E_\pi$ and the decay widths.
We find that $\hat g_{\rho\pi\pi}$ initially decreases quickly
with density and stabilizes beyond nuclear saturation density.

\begin{figure}[htb]
\centerline{\epsfig{file=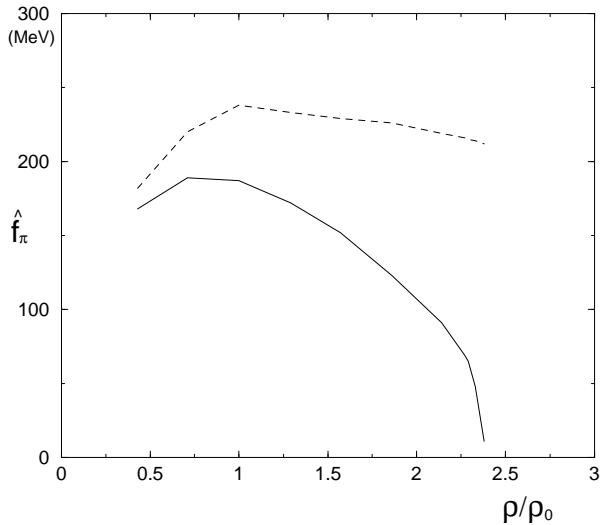,width=8cm}}
\vspace{0.3cm}
\caption{\small In-medium $\hat f_\pi$ in the R/W (dashed lines) and
B/R (solid lines) framework.}
\label{fph}
\end{figure}

We are now in position to assemble the main result of this section. 
Evaluating (\ref{relation}) with the additional input of the 
density-dependent $\hat\rho$ and $\hat a_1$ masses as found in 
Sect.~\ref{sec_sobar} (for the $\Gamma_{tot} =0$ case) results in an 
in-medium pion decay constant $\hat f_\pi(\rho)$ as shown in Fig.~\ref{fph}.
As anticipated from the left panel of Fig.~\ref{fig_schem}, 
the BR prescription leads to
a vanishing value at about 2.5$\rho_0$, whereas in the R/W
comparatively little density dependence is observed. However, as
mentioned before, the latter might still encode an approach
towards chiral restoration, albeit in a somewhat different fashion
which cannot be deduced from the behavior of the quasiparticle
masses. This issue will be addressed in Sect.~\ref{sec_wsr}.

%%%%%%%%%%%%%%%%%%%%%%%%%%%%%%%%%%%%%%%%%%%%%%%%%%%%%%%%%%%%%%%
\subsection{Relation to BR Scaling}
%%%%%%%%%%%%%%%%%%%%%%%%%%%%%%%%%%%%%%%%%%%%%%%%%%%%%%%%%%%%%%%
Let us now comment on the (plausible) connection between our sobar
field and effective fields of BR scaling. As was shown in 
Sect.~\ref{sec_schem}, the pole strength ($Z$-factor) of the 
sobar fields is less
than unity while the  effective fields of BR scaling carry full
pole strength. This discrepancy can be (partly) reconciled by a
finite wave-function renormalization (rescaling) of the
sobar-fields (which somewhat resembles the (infinite)
wave-function renormalization of free fields). Consider, \eg, the
$\hat\pi$-field. Expanding $\hat U$ in terms of pisobar fields,
one has \ba {\cal L}_{\hat\pi} = \frac{1}{2}\partial_{\mu}{
\hat\pi} \partial^\mu{\hat\pi} -\frac{1}{2}{m_\pi^2}{\hat\pi}^2
+c_1\frac{1}{{ \hat f}_\pi^2 }
(\partial_\mu\hat\pi)(\partial^\mu\hat\pi) \hat\pi\hat\pi
 +c_2 \frac{m_\pi^2}{\hat f_\pi^2} \hat\pi^4+...
\ea with coefficients $c_1$ and $c_2$. Since the corresponding
two-point function (propagator) is given by \ba
<\hat\pi\hat\pi^\dagger>\approx \frac{Z_{\hat\pi}}{k^2 -m_\pi^2},
\ea we rescale the field according to $\hat\pi =
Z_{\hat\pi}^{1/2}\pi^* $, so as to obtain a residue of $1$ at
$k^2= m_\pi^2$. This leads to \ba {\cal L}_{\hat\pi} =
\frac{1}{2}Z_{\hat\pi} \partial_\mu{\pi^*} \partial^\mu{\pi^*}
+c_1 \frac{1}{{\hat f}_\pi^2} Z_{\hat\pi}^2 (\partial_\mu \pi^*)
(\partial^\mu\pi^*)\pi^*\pi^* -\frac{1}{2}Z_{\hat\pi}
m_\pi^2{\pi^{*2} +c_2
 Z_{\hat\pi}^2 \frac{m_\pi^2}{{\hat f}_{\pi }^2}} \pi^{*4}+...~.
\ea
A redefinition of coupling constant and mass,
\ba
m_\pi^* = m_\pi {Z_{\hat\pi}}^{1/2},~
f_\pi^* ={\hat f}_\pi {Z_{\hat\pi}^{-1/2}},
\label{sc1}
\ea
then gives
\ba
{\cal L}_{\pi^*} = \frac{1}{2}Z_{\hat\pi} \partial_\mu\pi^* \partial^\mu\pi^* +c_1
Z_{\hat\pi}\frac{1}{{ f}_\pi^{*2}}
(\partial_\mu\pi^*)(\partial^\mu\pi^*) \pi^*\pi^*
-\frac{1}{2}m_\pi^{*2}\pi^{*2} +c_2 \frac{m_\pi^{*2}}{{ f}_\pi^{*2}}\pi^{*4} +...~.
\ea
A proper normalization of the  kinetic term can be recovered by
introducing a scale transformation
\ba
x\rightarrow x^\prime = Z_{\hat\pi}^{-1/2}x
\nonumber.
\ea
so that we obtain
\ba
{\cal L}_{\pi^*} = \frac{1}{2} \partial_\mu\pi^* \partial^\mu\pi^*
+c_1\frac{1}{{ f}_\pi^{*2}}
(\partial_\mu\pi^*)(\partial^\mu\pi^*) \pi^*\pi^*
-\frac{1}{2}m_\pi^{*2}\pi^{*2} +c_2 \frac{m_\pi^{*2}}{{ f}_\pi^{*2}}\pi^{*4}
+{\cal L}_{sb} + ...
\ea
However, since our sobar-Lagrangian cannot be expected to be fully
scale invariant (as opposed to a Lagrangian
exhibiting BR scaling\cite{br}), there appear both  
terms corresponding to those of BR scaling
as well as stemming from broken scale invariance (denoted by ${\cal L}_{sb}$).
Carrying out the same procedure for vector mesons, we find
\ba
m_\rho^*&=&Z_\rho^{1/2}{\hat m}_\rho,~ m_a^*=Z_a^{1/2}{\hat m}_a\no
 g_{\rho\pi\pi}^* &=& {\hat g}_{\rho\pi\pi} Z_\rho^{1/2} Z_{\hat\pi}^{1/2}.\label{sc2}
\ea
Using (\ref{sc1}),(\ref{sc2}) and neglecting $L_{sb}$, (\ref{relation})
finally becomes
\ba
 f_\pi^{*2} &\approx & \frac{2}{ g^{*2} }\frac{ m_\rho^{*2}
 }{ m_{a}^{*2} }( m_{a}^{*2} - m_\rho^{*2} )
\ea
where we have approximated
$Z_{a_1}\approx Z_\rho$ (cf.~right panel of Fig.~\ref{fig_schem}). Note that
$m_\pi^* = m_{\hat\pi}Z_{\hat\pi}^{1/2}$ does not
imply large changes for the in-medium pion mass, since
$Z_{\hat\pi}$ does not vary appreciably with density (see, \eg,
Ref.~\cite{mws}).

These observations indicate that the effective fields of
BR-scaling could be identified with a coherent linear combination
of the resonance-nucleon-hole degrees of freedom at some relevant
density. Similar arguments have been presented  in
Ref.~\cite{greiner} where the authors identified the sigma field
in the Walecka model with the degrees of freedom associated with
the $\Delta(1232)$ in the context of quantum hadrodynamics.
However, we repeat that -- as opposed to a Lagrangian leading to
BR scaling\cite{br} -- our effective meson-sobar Lagrangian is not
invariant under scale transformation symmetry. A possible
extension to restore scale invariance might be the introduction of
appropriate dilaton fields.

%%%%%%%%%%%%%%%%%%%%%%%%%%%%%%%%%%%%%%%%%%%%%%%%%%%%%%%%%%%%%%%%
\section{In-Medium Weinberg Sum Rules}
\label{sec_wsr}
%%%%%%%%%%%%%%%%%%%%%%%%%%%%%%%%%%%%%%%%%%%%%%%%%%%%%%%%%%%%%%%%
In this section we will pursue a somewhat different strategy to
assess the degree of chiral restoration realized within the schematic 
model of Sect.~\ref{sec_schem}. Rather than imposing the mean-field
approximation, we here account for the full spectral shape of both
$\rho$ and $a_1$ spectral functions. A rather direct way of
doing so is provided by the Weinberg sum rules~\cite{Wein67}, one
of which relates the pion decay constant (being one of the order
parameters of chiral restoration) to the (integrated) difference
in the spectral shapes between the vector ($\rho_V$) and
axialvector ($\rho_A$) correlators, \ie,
\be
f_\pi^2=\int\limits_0^\infty \frac{ds}{s} \left[\rho_V(s)-\rho_A(s)\right]
 \ .
\label{wsr1}
\ee
A second Weinberg sum rule applies to the next higher moment of the
correlator difference according to
\be
0=\int\limits_0^\infty ds \left[\rho_V(s)-\rho_A(s)\right]  \ .
\label{wsr2}
\ee
Whereas the first one, (\ref{wsr1}), is firmly established in its
given form with no known corrections, the second one, (\ref{wsr2}),
is likely to be modified away from the chiral limit~\cite{BDLW75}
and due to $U_A(1)$ symmetry breaking~\cite{Dm98}. From a practical 
point of view,
it is clear that within the scope of our low-energy effective model,
the first sum rule is the more relevant one.

In the low-mass region the correlators can be accurately
saturated within the (axial-) vector dominance model (VDM) by the
spectral function of the lowest lying resonances $\rho$ and $a_1$
according to
\be
\rho_{V,A}(q_0,q)= - \frac{(m_{V,A}^{(0)})^4}{\pi g_{V,A}^2}
                  \ {\rm Im} D_{V,A}(q_0,q)
\ee with $m_{V,A}^{(0)}$ denoting the bare pole masses and
$g_{V,A}$ the (axial-) vector coupling constants. However, the
$\rho$ and $a_1$ spectral functions account for two- and
three-pion states only. Therefore, at higher masses, the
contributions from $2n$- and $(2n+1)$-pion states (with $n>1$)
have to be included. This can be done by an appropriate
continuum ansatz characterized by an onset threshold and
(asymptotic)  plateau value governed by perturbative QCD,
\eg~\cite{Sh93},
\be
\rho_{V,A}^{cont}(s)=\frac{s}{8\pi^2}
\frac{1}{1+\exp\left[(E^{thr}_{V,A}-\sqrt{s})/\delta_{V,A}\right]}
\left(1+\frac{0.22}{\ln[1+\sqrt{s}/(0.2{\rm GeV})]}\right) 
\ee
with $\delta_{V}=\delta_A=0.2$~GeV and $E^{thr}_{V}\simeq
1.3$~GeV, $E^{thr}_{A}\simeq 1.45$~GeV. The merging of the
continuum plateaus into their perturbative form is crucial to
ensure the convergence of the sum rules (one-gluon exchanges do
not distinguish between vector and axialvector channels).

The in-medium  extensions of these sum rules have been derived in
Ref.~\cite{KS94}.  In the zero-momentum limit to which we restrict
ourselves here, one has
\be
{f_\pi^*}^2=\int\limits_0^\infty \frac{dq_0^2}{q_0^2}
\left[ \rho_V(q_0,\vec q=0)
-\rho_A(q_0,\vec q=0) \right]
\label{wsr1-med}
\ee
and an analogous expression for the second one.
The right-hand-side ({\it r.h.s.}) of (\ref{wsr1-med})
is now readily evaluated employing the $\rho$ and $a_1$ spectral
functions.  As off-shell properties are of potential relevance, we
calculate the free $\rho$ selfenergy  microscopically using
experimental data on the pion electromagnetic form factor and
$p$-wave $\pi\pi$ phase shifts~\cite{RCW}. Similarly, we obtain
the (imaginary part of the) free $a_1$ selfenergy from decays
into $\rho\pi$ in terms of the free $\rho$ spectral
function~\cite{RCW,RG99} (where possible energy dependencies of
the real part are neglected, that is, we take
$(m_{a_1}^{(0)})^2+{\rm Re}\Sigma_{a_1}(s)\equiv m_{a_1}^2$; the
remaining free parameter, $g_{a_1}=7.8$, has been fixed via
 (\ref{wsr1}) by requiring $f_\pi\simeq 93$~MeV). The in-medium
resonance-hole selfenergies are then introduced as described in
Sect.~\ref{sec_schem}, \ie, (\ref{RWself}) and the analogous
expression for the $a_1N^*(1900)N^{-1}$ excitation (with
microscopic free widths for both $N^*(1520)$, $N^*(1900)$, and
including in-medium broadening).

\begin{figure}[htb]
\vspace{-0.5cm}
\centerline{\epsfig{file=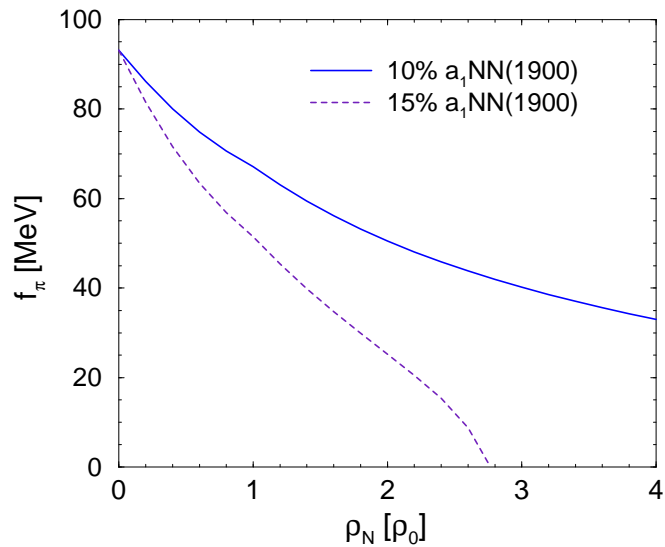,width=8cm,angle=-90}}
\vspace{0.5cm}
\caption{\small Density dependence of $f_\pi$ as extracted from
the first Weinberg sum rule using in-medium
$\rho$ and $a_1$ spectral functions within the schematic two-level
model; the two curves correspond to branching ratios of
10\% (solid line) and 15\% (dashed line) for the $N^*(1900)\to
Na_1$ decay.} \label{fig_wsr}
\end{figure}

\begin{figure}[b]
\vspace{-1cm}
\epsfig{file=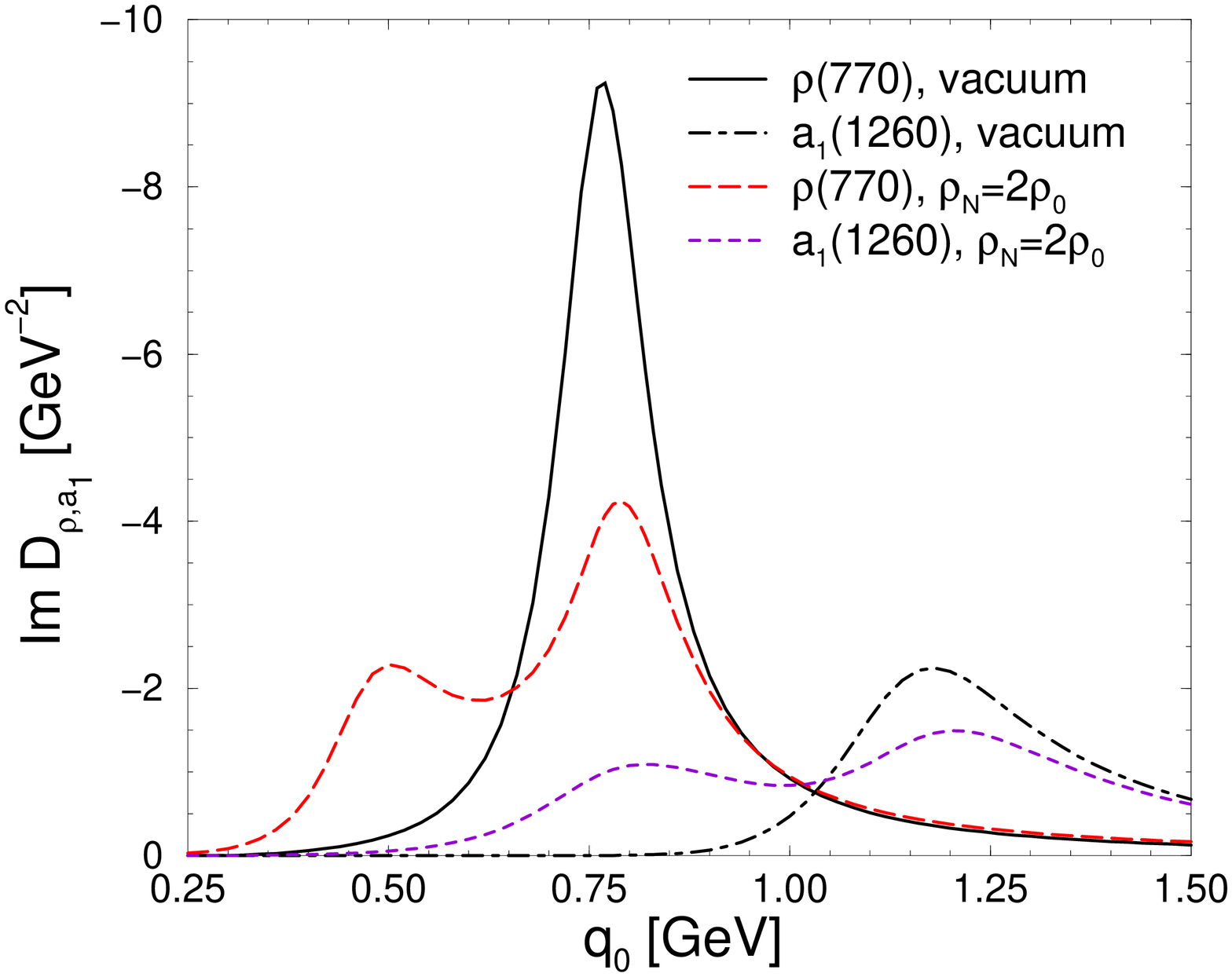,width=7.2cm,angle=-90}
\epsfig{file=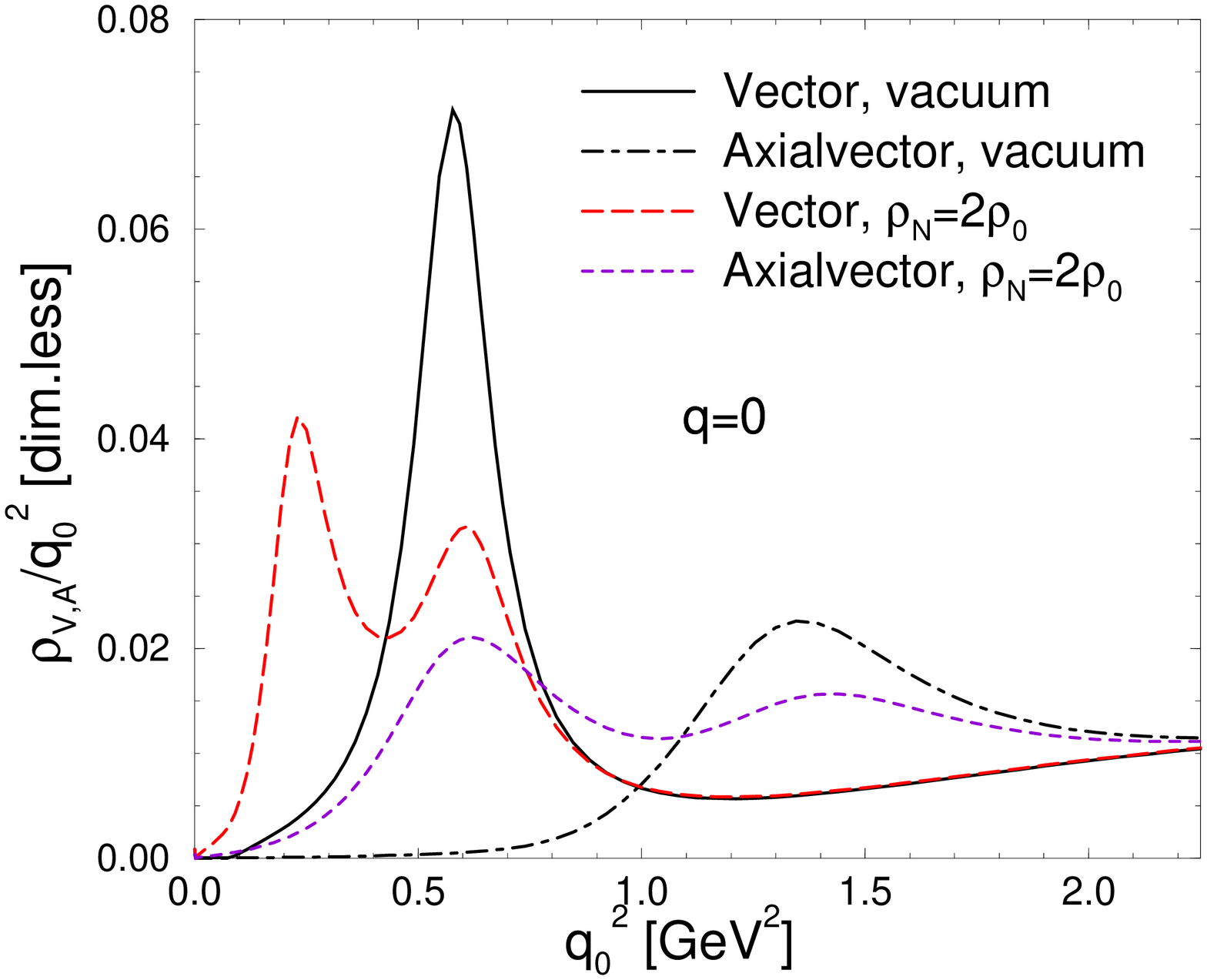,width=7.2cm,angle=-90}
\vspace{0.5cm}
\caption{\small $\rho$ and $a_1$ spectral functions (left panel)
and $V$/$A$ correlators divided by $q_0^2$ (right panel)
within the schematic resonance-hole model; full lines: free $\rho$ ($V$),
dotted lines: free $a_1$ ($A$), long-dashed and dashed lines: $\rho$ ($V$)
and $a_1$ ($A$) at $\rho_N=2\rho_0$.}
\label{fig_spec}
\end{figure}

Fig.~\ref{fig_wsr} shows that the pion decay constant indeed
decreases rather rapidly starting from low densities reaching a
30-40\%  reduction at normal nuclear density, depending on the
precise value for the $a_1N^*(1900)N$ coupling constant. The
uncertainty quickly increases towards higher densities, resulting
in either a leveling-off or even a dropping to zero. However, the
point here is not so much to predict quantitatively the critical
density for chiral restoration but to demonstrate {\it
qualitatively} that in-medium spectral functions based on
many-body excitations in the nuclear environment encode an 
approach towards chiral restoration. 
To gain  more insight into
the underlying mechanism we show in Fig.~\ref{fig_spec} the
pertinent vector and axialvector spectral distributions in vacuum
and at twice normal nuclear density: clearly, the chiral breaking
in vacuum, mainly constituted through different $\rho$ ans $a_1$
pole positions, is much reduced in the medium due to
the appearance of low-lying excitations as well as a smearing of
the elementary $\rho$/$a_1$ peaks.  The schematic two-level model
employed here certainly lacks accuracy, but it is conceivable that
the inclusion of further excitations will reinforce the tendency
towards degeneration of the spectral densities.

%%%%%%%%%%%%%%%%%%%%%%%%%%%%%%%%%%%%%%%%%%%%%%%%%%%%%%%%%%%%%%%%
\section{Summary and Conclusion}
\label{sec_concl}
%%%%%%%%%%%%%%%%%%%%%%%%%%%%%%%%%%%%%%%%%%%%%%%%%%%%%%%%%%%%%%%%
The identification of experimental signatures as well as
theoretical mechanisms associated with chiral symmetry restoration
in hot/dense matter has become an important issue in
(nonperturbative) strong interaction physics. In this article we
have focused on the behavior of the vector and axialvector
correlator in cold nuclear matter which have to degenerate at the
critical density. Using phenomenologically well-motivated --
albeit somewhat schematic -- (two-level) models based on low-lying
resonance-hole ('sobar') excitations, we have pursued two
approaches to assess possible avenues towards chiral restoration.
In addition to the well-established $\pi$-sobar and $\rho$-sobar
states, we have furthermore included the $a_1$-sobar based on the
$N(1900)N^{-1}$ excitation as a candidate for the chiral partner
of the $N(1520)N^{-1}$ state.

In the first part, we constructed a chiral effective Lagrangian
for the `meson-sobar' fields supposed to be valid in the medium,
incorporating vector and axialvector degrees of freedom within
the Massive Yang-Mills framework (MYM). The parameters entering  
this Lagrangian have been estimated from the schematic model.
Within a mean-field approximation, and imposing   
a general matching condition inherent in the MYM formalism,  
we found that the in-medium gauge coupling $\hat g$ and the (axial-) 
vector meson masses decrease with density. Performing a rescaling of 
the meson-sobar fields, we argued that the latter might be identified
with the fields of BR scaling. {\it Our proposition is that these
are the degrees of freedom that figure in effective Lagrangian
field theory endowed with chiral symmetry and other flavor
symmetries and that are probed in the dilepton production
experiments at high density.} The fields that figure in BR scaling
are therefore collective fields carrying the quantum numbers of
mesons and baryons treated as local fields and hence must
necessarily be approximate, with their range of validity depending
on kinematics.
%(we should note, however,  that our effective MYM Lagragian of meson-sobar
% fields is not strictly scale invariant as opposed to within the BR case)

In the second part we have made use of the in-medium extension of
the first Weinberg sum rule to evaluate the density dependence of
the pion decay constant. Employing in-medium many-body spectral
functions for both the $\rho$ and $a_1$ as arising from the
two-level model, together with appropriate (density-independent)
high energy continua, $f_\pi(\rho_N)$ exhibits an appreciable
reduction of $\sim 30$~\% at normal nuclear matter density. The
driving mechanism in this calculation has been identified as a
growing overlap of the $V$- and $A$-spectral distributions due to
both the appearance of the low-lying `sobar'-excitations as well
as an accompanying resonance broadening.

Although our modeling of the various effective interactions at
finite density is far from complete, we expect that the qualitative
features of our results will persist in a more elaborate
treatment. The latter might also provide a better estimate for the
critical density with most of the uncertainty likely to reside in
the $a_1$ channel where there is little empirical information
available. QCD-based models (\eg, with  instanton or
NJL-type interactions) typically give a rather low value for
$\rho_c$ (around $2\rho_0$), which might well be due to the
neglect of interactions at the composite (hadronic) level. So far
these are more reliably addressed in effective hadronic models as
the one presented here.

\section*{ACKNOWLEDGMENTS}
We thank C.-H. Lee and J. Wambach for useful discussions. This
work was supported in part by DOE grant No. DE-FG02-88ER40388. YK
is supported in part by the Korea Ministry of Education (BSRI
99-2441) and KOSEF (Grant No. 985-0200-001-2).

%%%%%%%%%%%%%%%%%%%%%%%%%%%%%%%%%%%%%%%%%%%%%%%%%%%%%%%%%%%%%%%%%%%%%

\end{document}